\documentclass[letter,twocolumn]{jpsj3}
\usepackage{txfonts}
\usepackage{amsmath}
\usepackage{bm}
\usepackage{color}

\title{Onsager vortex formation in two-component Bose--Einstein condensates}

\author{Junsik Han$^1$, and Makoto Tsubota$^{1,2}$}
\inst{$^1$Department of Physics, Osaka City University, 3-3-138 Sugimoto, Sumiyoshi-ku, Osaka 558-8585, Japan \\
        $^2$The Advanced Research Institute for Natural Science and Technology, Osaka City University, 3-3-138 Sugimoto, Sumiyoshi-ku, Osaka 558-8585, Japan}
\date{\today}

\abst{We numerically study the dynamics of quantized vortices in two-dimensional two-component Bose--Einstein condensates (BECs) trapped by a box potential.
For one-component BECs in a box potential, it is known that quantized vortices form Onsager vortices, which are clusters of same-sign vortices.
We confirm that the vortices of the two components spatially separate from each other--even for miscible two-component BECs--suppressing the formation of Onsager vortices. 
This phenomenon is caused by the repulsive interaction between vortices belonging to different components, hence, suggesting a new possibility for vortex phase separation.
}

\begin{document}
\maketitle


In turbulence, vortices create high concentrations of vorticity ${\bm \omega}({\bm r}, t) = {\bm \nabla} \times {\bm v}({\bm r}, t)$, where ${\bm v}({\bm r}, t)$ is velocity field of the fluid.
Turbulence and the related vortex behavior exhibit universal statistical laws, which depend on the spatial dimensions of the system \cite{Davidson15}.
Three-dimensional (3D) turbulence sustains a direct cascade of energy transfer from the larger to smaller spatial scales.
This direct cascade is believed to be sustained by the Richardson cascade, in which large vortices are broken up into smaller ones.
On the other hand, the behavior of two-dimensional (2D) turbulences is in stark contrast with 3D turbulences.
Onsager predicted the spontaneous formation of large-scale, long-lived vortices which are called Onsager vortices \cite{Onsager49,Eyink06}.
Kraichnan predicted that 2D turbulence exhibits an inverse energy cascade from the smaller to larger spatial scales \cite{Kraichnan67}.

When we try to study Richardson cascades, classical turbulence (CT) causes serious difficulties due to the difficulty of defining each vortex.
Quantum turbulence (QT) has the advantage of considering quantized vortices as topological defects.
This means that vortices are well-defined in both 3D and 2D systems for QT.
An atomic Bose--Einstein condensate (BEC) is a typical system of a quantum fluid and has the following advantages for the study of turbulence.
Firstly, the diluteness of BEC gas makes the vortex core relatively large, reaching values in the $\mu {\rm m}$ range, so that the vortex cores are visible by optical techniques.
Secondly, BEC can be controlled well experimentally, for example, the intensity of the atom-atom interactions can be changed through Feshbach resonance.
Finally, we can use the Gross-Pitaevskii (GP) equation--based on the mean field approximation--to treat the dynamics of the condensate,
which describes the experimental results quantitatively.

Three-dimensional QT also sustains direct energy cascades and exhibits Kolmogorov's $-$5/3 power law, which may support the idea that the Richardson cascade process is present in the system \cite{Kobayashi05}.
From this similarity between the QT and CT in 3D systems, we can expect the formation of Onsager vortices in 2D QTs.
Therefore, several experimental, numerical, and theoretical studies have been done, aimed to prove whether or not Onsager vortices are formed in 2D QTs \cite{Neely13,Seo17,Gauthier18,Johnstone18,Numasato10,White12,Reeves13,Billam14,Simula14,Billam15,Groszek16,Yu16}.
Specifically, Groszek  {\it et al.} showed that the formation of Onsager vortices depends on the trapping potential by carrying out simulations based on the GP model.
In order to provide a definite proof, they calculated the amplitude of the dipole moment of the vortex distribution defined by $d=|{\bm d}|=|\Sigma_{i}q_{i}{\bm r}_{i}|$, where ${\bm r}_{i}$ is the position of the $i$th vortex and $q_{i} = s_{i}\kappa = s_{i}h/m$ is its charge with $s_{i}=\pm1$. 
They concluded that Onsager vortices are formed in uniform condensates trapped by a box potential \cite{Gaunt13,Navon16}.

These studies addressed one-component BECs.
On the contrary, two-component BECs were studied in 2D \cite{Kasamatsu03,Schweikhard04,Eto11,Nakamure12,Karl132D,Kasamatsu16} and in 3D \cite{Takeuchi10,Karl133D}, observing novel phenomena not found in one-component BECs.
For example, vortices form interlocked triangular lattices, square lattices, or interwoven serpentine vortex sheets in 2D rotating condensates.
These phenomena originate from the intracomponent- and intercomponent-coupling of the condensates.
The intracomponent-coupling of the condensates is the interaction between atoms of the same condensate component denoted by $g_{ii} \ (i = 1,2)$. The intercomponent-coupling of the condensates is the interaction between atoms of different condensate components denoted by $g_{12}$.
Here, two components are miscible when $\sqrt{g_{11}g_{22}} > g_{12}$ and phase separated when $\sqrt{g_{11}g_{22}} < g_{12}$.

These two coupling types of the condensates result in two kinds of vortex interactions: intracomponent- and intercomponent-interaction.
Intracomponent-interaction of the vortices is the interaction between vortices belonging to the same condensate component, while intercomponent-interaction of the vortices means the interaction between vortices two different components.
The energy of the intracomponent-interaction of the vortices is written by 
\begin{equation}
\epsilon^{\rm intra}_{ij} = \frac{2\pi s_{i} s_{j} \hbar^{2} n}{m} \ln \frac{R_{0}}{R_{ij'}}, \label{intracomponent-interaction}
\end{equation}
where $n$ is the density of the condensates, $m$ is the mass of the involved atom, $R_{ij}$ is the distance between the $i$th and $j$th vortices and $R_{0}$ is the radius of the potential. \cite{Pethick-Smith}.
Then, $s_{i}$ and $s_{j}$ are the circulation signs of the $i$th and $j$th vortices, respectively. If these circulation signs are equal, the intracomponent-interaction is repulsive, and attractive if the signs differ.
The energy of the intercomponent-interaction of vortices is 
\begin{equation}
\epsilon^{\rm inter}_{ij} = \frac{\pi \hbar^{4} g_{12}}{4 m_{1} m_{2} (g_{11}g_{22} - g^{2}_{12})} \frac{\ln \frac{R_{ij}}{\xi}}{R^{2}_{ij}}, \label{intercomponent-interaction}
\end{equation} 
where $m_{i}$ is the mass of the $i$th component's atom, $2R_{ij}$ is the distance between the $i$th and $j$th vortices belonging to two different components, and $\xi$ is the coherence length.\cite{Eto11}.
Then, whether this interaction is repulsive or attractive depends on the signs of $g_{12}$ and $g_{11}g_{22} - g^{2}_{12}$.
The two interactions show a different dependence on the distance between vortices.

The presence of the two interactions causes the vortex behaviors of two-component BECs more complex than that present in one-component BECs.
Here, we study a two-component BEC system.
We expect that the formation of Onsager vortices will depend on the competition between the two vortex interactions.
The main interest of this work is to elucidate if Onsager vortices form and how the vortex dynamics are affected by the intercomponent-interactions. 
Hence, we consider BECs trapped by a box potential, which was also realized experimentally \cite{Gaunt13,Navon16}, as the formation of Onsager vortices was confirmed for a one-component BEC within the same conditions \cite{Groszek16}.


In this study, we address 2D two-component BECs trapped by a box potential.
The $i$th component of the Bose--Einstein condensate can be described by a macroscopic wave function $\psi_{i} = \sqrt{n_{i}({\bm r},t)}e^{\imath \phi_{i}({\bm r},t)}$ where $n_{i}({\bm r},t)$ is the density of the condensates and $\phi_{i}({\bm r},t)$ is its phase.
These wave functions obey the GP equations
\begin{equation}
\begin{array}{l}
 \imath\hbar\frac{\partial}{\partial t}\psi_{i}({\bm r},t) \\
 \\
 = \left[\frac{-\hbar^{2}}{2m_{i}}\nabla^{2} + V_{\rm trap}({\bm r}) + \sum_{j=1,2} g_{ij}|\psi_{j}({\bm r},t)|^{2}\right] \psi_{i}({\bm r},t), \\
 \\
   \hspace{1pc}(i = 1, 2)\label{eq:GPEs}
\end{array}
\end{equation}
where $m_{i}$ is the mass of the $i$th component's atom. 
Here, we initially choose $m_{1} = m_{2} = m$, $g_{11} = g_{22} = g$, and $g_{12} = 0.1 g > 0$.
In order to focus on the dependence of the Onsager vortex formation on $g_{12}$, we consider two conditions: (i) $g_{12} = 0.1 g$ is constant; and (ii) we change $g_{12}$ from $0.1 g$ to $0.7 g$ at $t = 250$.
With condition (i), $d$ of both components increases steadily from $t = 0$ to $t \simeq 250$ (Fig. \ref{fig:2C_dipole_11}) and Onsager vortex formation is confirmed.
Then, we change $g_{12}$ from $0.1 g$ to $0.7 g$ at $t = 250$--according to condition (ii)--in order to determine the effect of $g_{12}$ on the Onsager vortex formation, namely, to what extent does $g_{12}$ affect the Onsager vortices.
With both conditions (i) and (ii), the two components are miscible.
We consider a box potential:
\begin{equation}
V_{\rm trap}({\bm r}) = \left \{
\begin{array}{l}
V_{0} \hspace{1pc}(|{\bm r}| > R_{0}) \\
0 \hspace{1.5pc}(|{\bm r}| < R_{0})
\end{array} 
\right.,\label{eq:box_potential}
\end{equation}
with a potential radius $R_{0}$ and potential height $V_{0}$.

To create the initial state, we imprint vortices in the condensate by multiplying the wave function by a phase factor $\Pi^{N_{v}}_{i}\exp(\imath \phi_{i})$, with $\phi_{i}(x,y) = s_{i}\arctan[(y-y_{i})/(x-x_{i})]$. 
Here, the coordinates $(x_{i},y_{i})$ refer to the position of the $i$th vortex and they are chosen randomly.
After the imprinting, the wave function evolves in imaginary time to establish the structure of the vortex cores.
We treat the state with the vortex cores formed as an initial state.
Subsequently, we solve Eq. (\ref{eq:GPEs}) in real time using the Fourier and Runge-Kutta methods on a $512 \times 512$ spatial grid.

The vortices are identified by finding the phase singularities of the wave function. 
The number of vortices is counted in a region $|{\bm r}| < 0.9 R_{0}$ in order to avoid counting ghost vortices in the low-density region \cite{Tsubota02}.
We also calculate the amplitude of the dipole moment of the vortex distribution--which is defined as $d=|{\bm d}|=|\Sigma_{i}q_{i}{\bm r}_{i}|$--to characterize the formation of Onsager vortices. 
If the vortices are distributed uniformly, $d$ seldom grows. 
If like-sign vortices form an Onsager vortex, $d$ develops into a finite value.


We numerically modeled the formation of Onsager vortices in one-component BECs by using GP, and obtained results consistent with the presented in a previous study \cite{Groszek16}.
Then, we extended the system to contain two components.

\begin{center}
\begin{figure}[h]
\includegraphics[width=20pc, height=11.5pc]{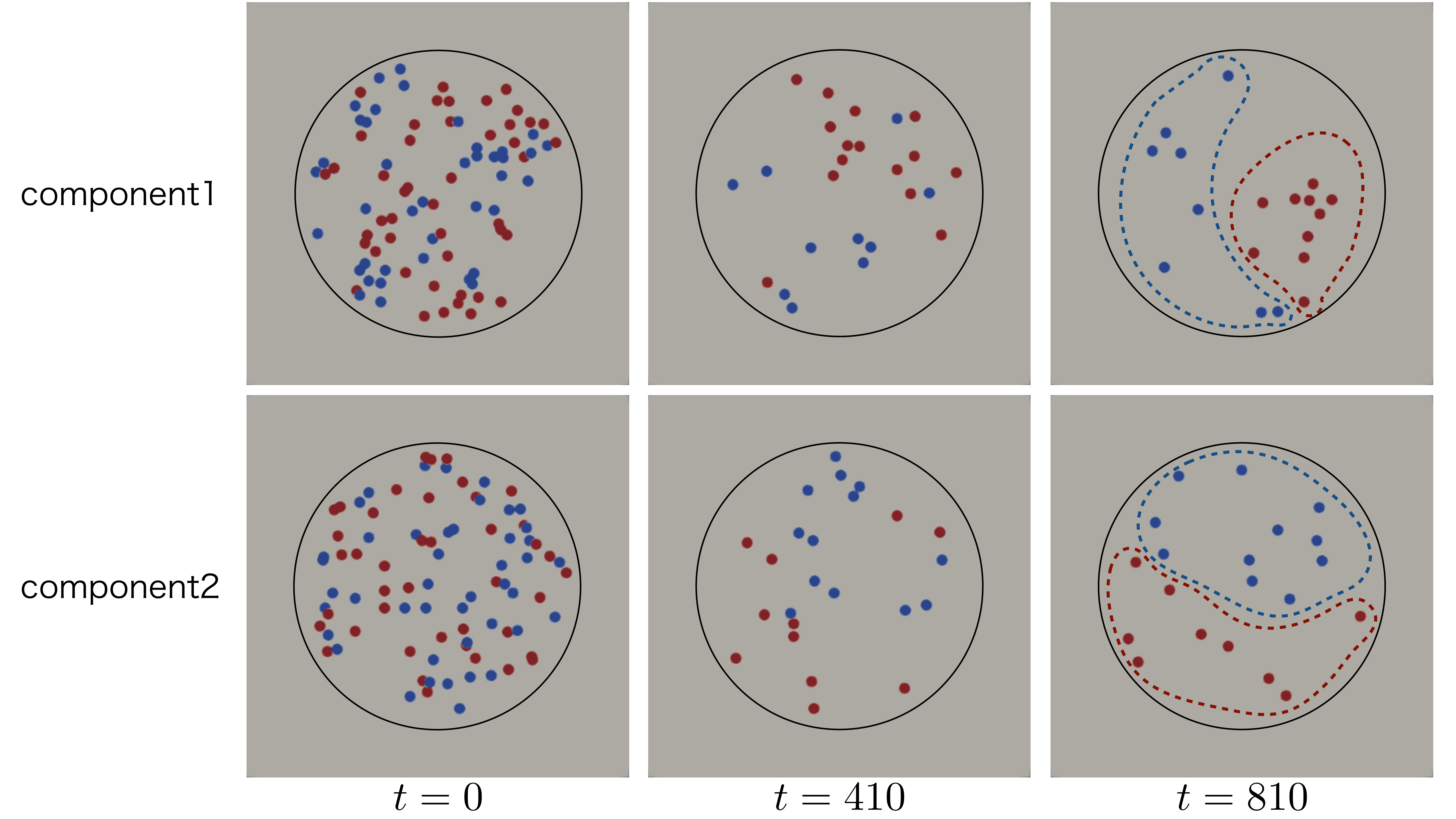}
\caption{\label{fig:2C_point_11}The vortex distribution in a two-component BEC at $t = 0,\  410,\   810$.
The intercomponent-coupling $g_{12} = 0.1g$ is constant.
Vortices and antivortices are denoted by red and blue points, respectively. Black circles represent the boundary of the condensate. 
The dotted red and blue lines surround the clusters of vortices and antivortices, respectively.}
\end{figure}
\end{center}

\begin{center}
\begin{figure}[h]
\includegraphics[width=21pc, height=10pc]{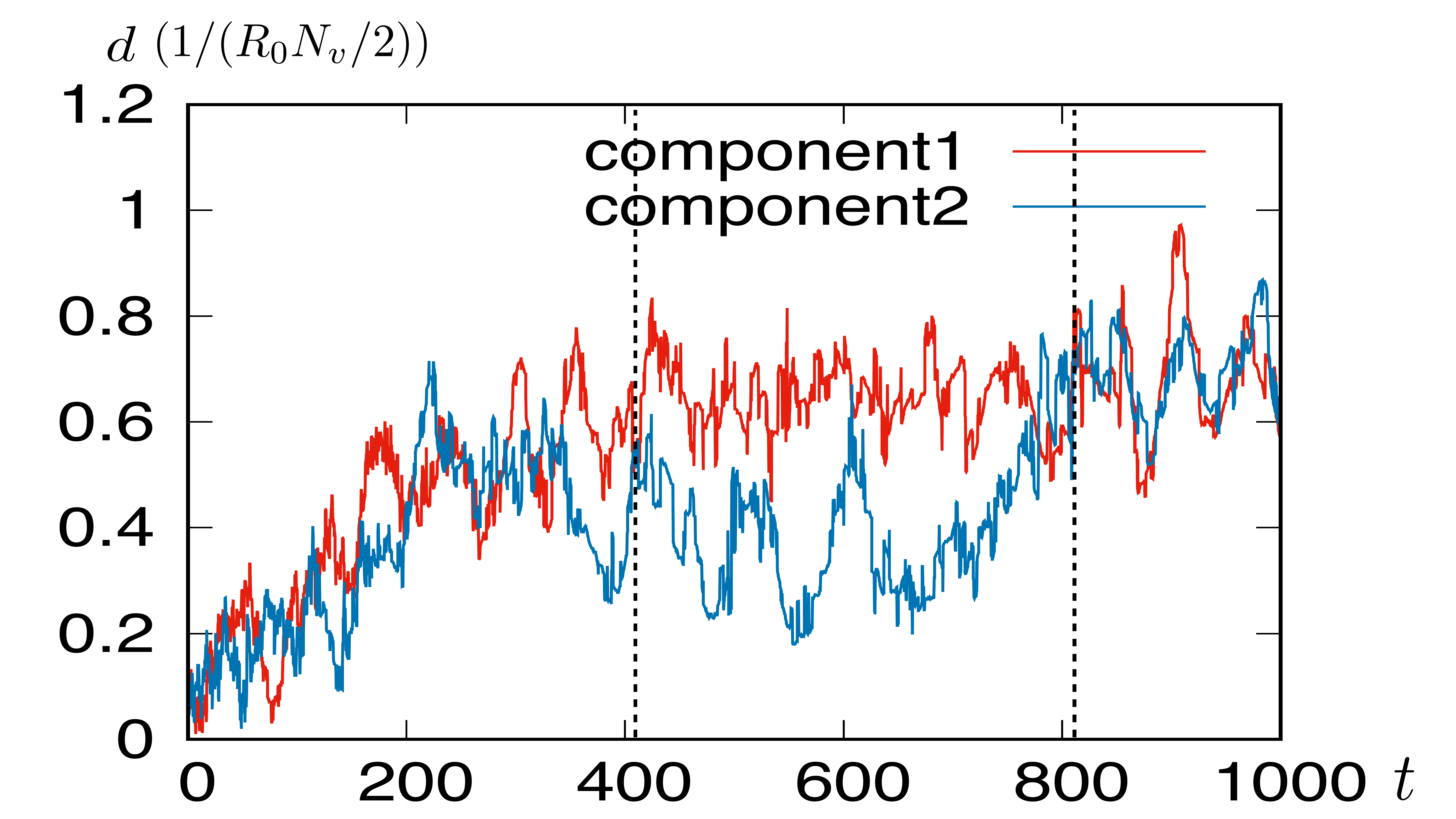}
\caption{\label{fig:2C_dipole_11}Time development of the dipole moment $d$ in a two-component BEC.
The intercomponent-coupling $g_{12} = 0.1g$ is constant.
The dipole moment is normalized by the radius of the potential ($R_{0}$), vortex charge ($\kappa$), and vortex pair number ($N_{v}/2$).
The two vertical dotted lines correspond to the time frames displayed in Fig. \ref{fig:2C_point_11}}
\end{figure}
\end{center}

We first focus on the result with condition (i).
Figure \ref{fig:2C_point_11} shows temporal evolution of the distribution of the vortices, while Fig. \ref{fig:2C_dipole_11} shows the time development of $d$.
The initially uniform vortices lead to two Onsager vortices: one consisting of vortices ($s_{i} = 1$) and the other consisting of antivortices ($s_{i} = -1$) at $t = 810$ in both components (Fig. \ref{fig:2C_point_11}). 
Here, decay of the number of vortices results from vortex-antivortex annihilation \cite{Groszek16}.
The $d$ in both components increases to finite values (Fig. \ref{fig:2C_dipole_11}). 
These results of each component are similar to the previous simulation \cite{Groszek16} considering one-component BECs. 
This can be attributed to the following reason:
if $g_{12} = 0$, each component is independent and vortices belonging to one component are independent of the vortices belonging to the other component. Hence, their states are perfectly equivalent to that of a one-component BEC.
If $g_{12}$ is small, the intercomponent-interaction of the vortices is weak, leading to the states of each component being similar to that of one-component BECs. 

\begin{center}
\begin{figure}[h]
\includegraphics[width=20pc, height=11.5pc]{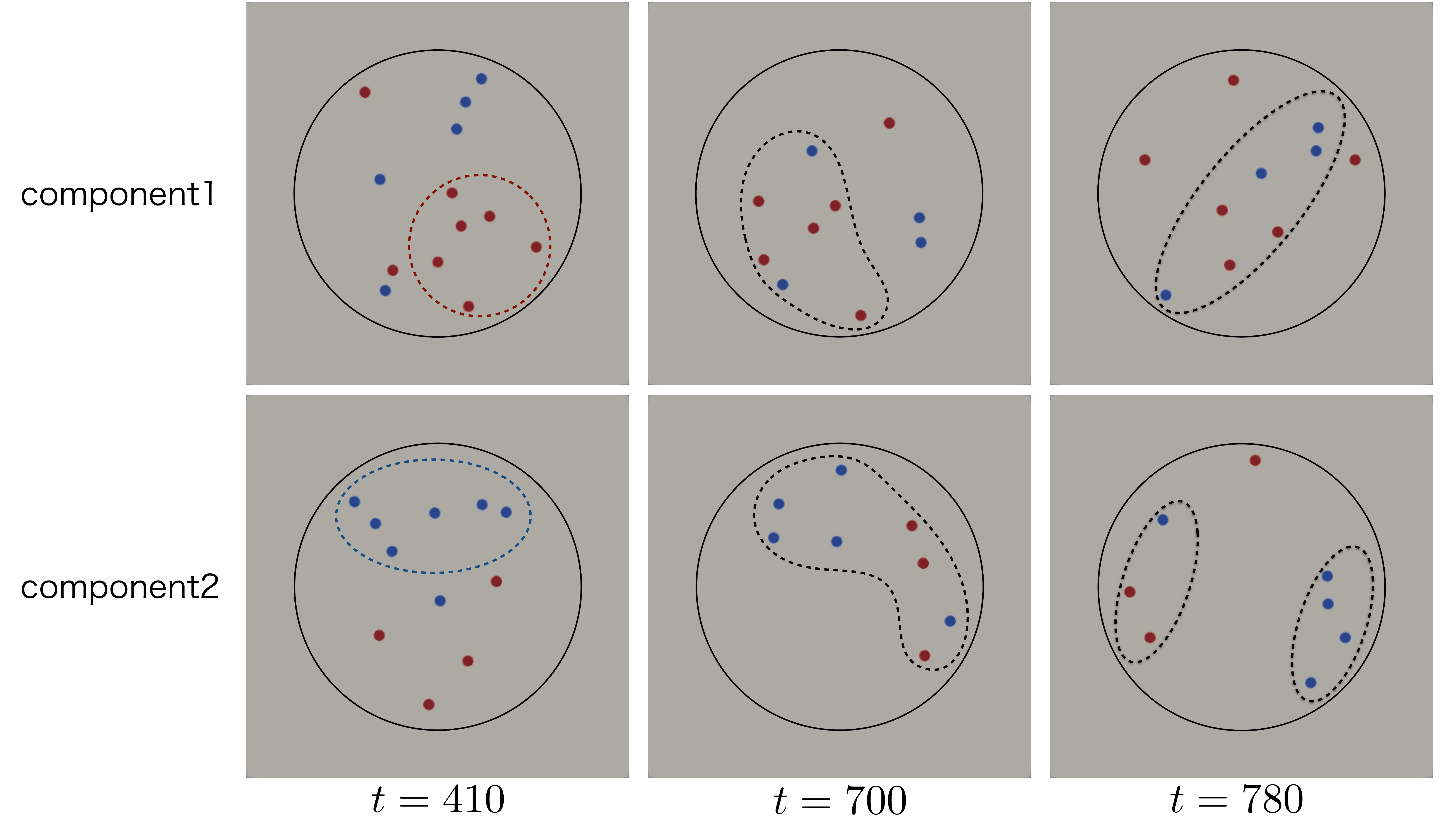}
\caption{\label{fig:2C_point_17}The distribution of vortices in a two-component BEC at $t = 410,\  700,\  780$.
The initial intercomponent-coupling $g_{12} = 0.1g$ is changed to $g_{12} = 0.7g$ at $t = 250$.
Vortices and antivortices are denoted by red and blue points, respectively. The black circles represent the boundary of the condensate. 
The dotted red and blue lines surround the cluster of vortices and antivortices, respectively.
At $t = 700$ and $t =780$, the vortices cluster in the region surrounded by the dotted black line in each component.}
\end{figure}
\end{center}

\begin{center}
\begin{figure}[h]
\includegraphics[width=21pc, height=10pc]{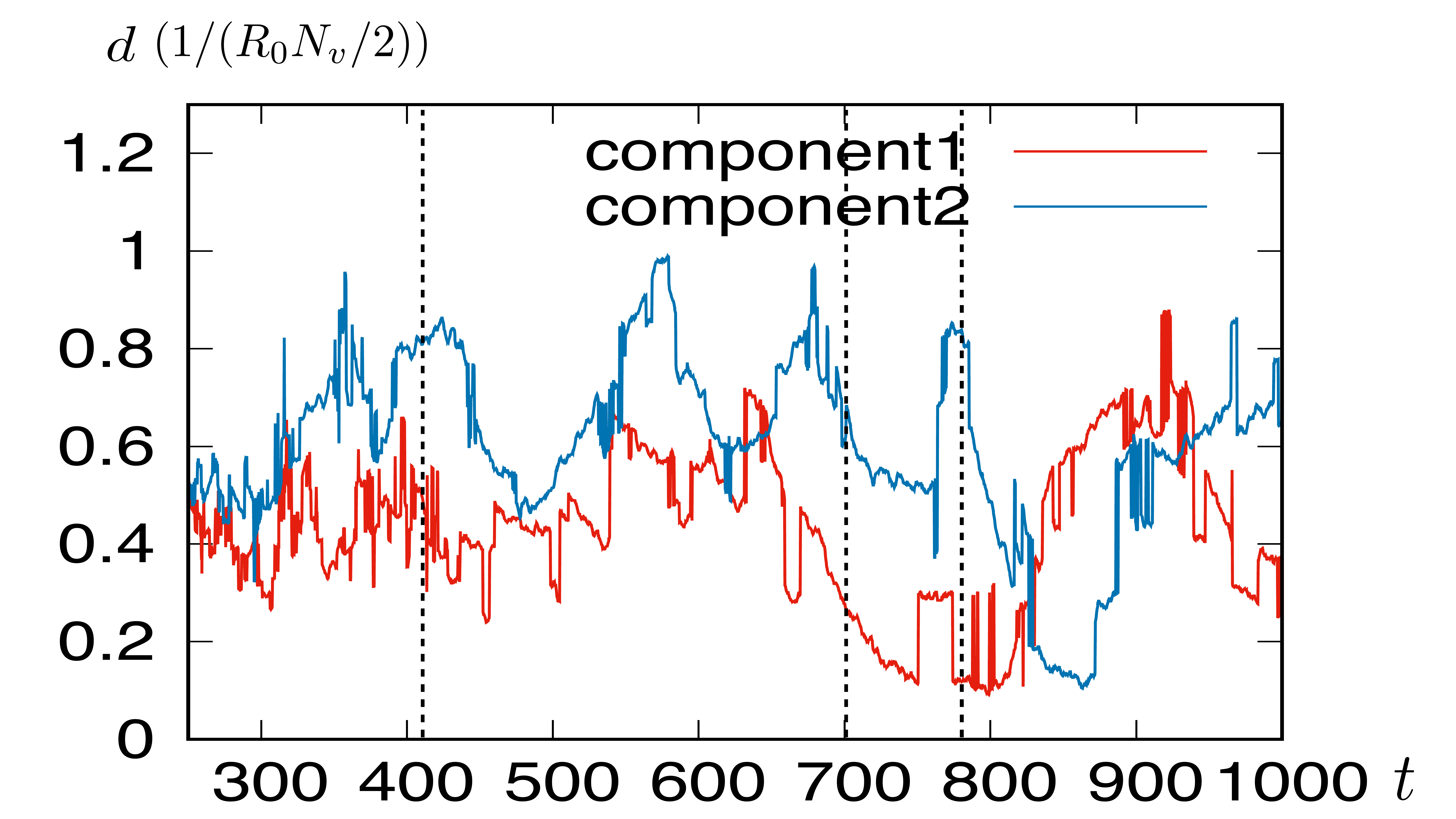}
\caption{\label{fig:2C_dipole_17}Time development of the dipole moment $d$ in a two-component BEC.
The initial intercomponent-coupling $g_{12} = 0.1g$ is changed to $g_{12} = 0.7g$ at $t = 250$.
The dipole moment is normalized by the radius of potential ($R_{0}$), vortex charge ($\kappa$), and vortex pair number ($N_{v}/2$).
The three vertical dotted lines correspond to the time frames displayed in Fig. \ref{fig:2C_point_17}.}
\end{figure}
\end{center}

With condition (ii), a novel phenomenon appears.
Figure \ref{fig:2C_point_17} shows the temporal evolution of the vortex distribution, while Fig. \ref{fig:2C_dipole_17} shows the time development of the $d$.
Here, the results from $t = 0$ to $t = 250$ are equal to the results of condition (i).

After $t \simeq 410$ (Fig. \ref{fig:2C_dipole_17}, left vertical dotted line), the time development of $d$ changes from that preceding $t \simeq 410$ qualitatively, which results from the phase separation of the vortex distribution.
Until $t \simeq 410$, $d$ of each component does not change significantly.
Then, we can confirm the tendency of the formation of Onsager vortices in each component (Fig. \ref{fig:2C_point_17}, left column, dotted red and blue lines).
However, after $t \simeq 410$, $d$ of both components vibrate with a lower frequency than that preceding $t \simeq 410$.
This time development of $d$ after $t \simeq 410$ results from the behavior of the region that the vortices occupy, which we call the vortex-region (VR). 
Around $t = 700$ and $t = 780$, the typical influence of the VR on the $d$ is shown (Fig. \ref{fig:2C_dipole_17} center and right vertical dotted lines).
The $d$ of both components decreases considerably around $t=700$.
At this time, the vortices in one component exhibit phase separation from those of the other component.
This means that the VR in each component becomes smaller than the whole area of the box potential, resulting in a significant decrease of $d$.

Around $t = 780$, the $d$ of component 1 decreases, while the $d$ of component 2 grows.
At $t = 780$, the vortices cluster in the region surrounded by the dotted black line for each component. Subsequently, the VR of component 2 is blocked by the VR of component 1, and is divided into two small VRs (Fig. \ref{fig:2C_point_17}, right column).
These vortex distributions reduce $d$ of component 1, and increase $d$ of component 2. 
As we can see at $t = 700$ and $t = 780$, we cannot confirm the formation of Onsager vortices, as with condition (i) after a sufficient time.
Around $t \simeq 420$, $570$, and $680$, we confirm the same tendency of the VR of component 2 being blocked and divided by the VR of component 1. This increases $d$ of component 2.
This phenomenon of phase separation is generally exhibited by this system.

We numerically calculated the formation of Onsager vortices in a two-component BEC trapped by a box potential, observing a novel phenomenon.
The VRs of each component become smaller than the whole size of the box potential through the effect of the intercomponent-interaction of vortices which results from the intercomponent-coupling of the condensates.
This phenomenon depends on the intensity of the latter.
For a weak intercomponent-coupling of the condensates, this phenomenon is not clearly observable.
On the contrary, for a strong intercomponent-coupling of the condensates, the phenomenon appears clearly. It suppresses the formation of the Onsager vortices, which spread over the whole region inside the potential trap, similarly to the weak coupling.

This phenomenon can be treated as the phase separation of the vortex distributions. 
The presence of this phenomenon is interesting because the vortex distributions separate even though the two components are miscible. In addition, it suppresses the formation of large-scale spatial structures, which were confirmed in one-component BECs.
We performed several equivalent runs for different initial vortex configurations, and obtained the qualitatively same results.

We will quantitatively study the details of the dependence of the dynamics and distributions of the vortices on the intensity of the intercomponent-coupling of the condensates in a follow-up paper.

\section*{Acknowledgment} \label{sec:Conclusion}
This  work  was  supported  by  JSPS  KAKENHI  Grant  No.   17K05548 and MEXT KAKENHI “Fluctuation \& Structure” Grant No. 16H00807

\bibliographystyle{jpsj}
\bibliography{book,aps,other}

\end{document}